\documentclass[conference,a4paper]{IEEEtran}
\IEEEoverridecommandlockouts

\usepackage{fancyhdr}
\usepackage{graphicx}
\usepackage{indent}
\usepackage{amsmath,amssymb}
\usepackage{paralist}
\usepackage{eclbkbox}
\usepackage{subfigure}
\usepackage{multirow}

\begin{document}
\title{Affective Recommendation System for Tourists\\by Using Emotion Generating Calculations
\thanks{\copyright 2014 IEEE. Personal use of this material is permitted. Permission from IEEE must be obtained for all other uses, in any current or future media, including reprinting/republishing this material for advertising or promotional purposes, creating new collective works, for resale or redistribution to servers or lists, or reuse of any copyrighted component of this work in other works.}
}

\author{\IEEEauthorblockN{Takumi Ichimura}
\IEEEauthorblockA{Department of Management and Systems,\\ Prefecture University of Hiroshima,\\ Hiroshima, 734-8558 Japan\\
E-mail: ichimura@pu-hiroshima.ac.jp}
\and
\IEEEauthorblockN{Issei Tachibana}
\IEEEauthorblockA{Graduate School of Comprehensive Scientific Research,\\
Prefectural University of Hiroshima\\
1-1-71, Ujina-Higashi, Minami-ku,\\
Hiroshima, 734-8559, Japan\\
Email: isseing1224@gmail.com}
}

\maketitle

\pagestyle{fancy}{
\fancyhf{}
\fancyfoot[R]{}}
\renewcommand{\headrulewidth}{0pt}
\renewcommand{\footrulewidth}{0pt}

\begin{abstract}
An emotion orientated intelligent interface consists of Emotion Generating Calculations (EGC) and Mental State Transition Network (MSTN). We have developed the Android EGC application software which the agent works to evaluate the feelings in the conversation. In this paper, we develop the tourist information system which can estimate the user's feelings at the sightseeing spot. The system can recommend the sightseeing spot and the local food corresponded to the user's feeling. The system calculates the recommendation list by the estimate function which consists of Google search results, the important degree of a term at the sightseeing website, and the the aroused emotion by EGC. In order to show the effectiveness, this paper describes the experimental results for some situations during Hiroshima sightseeing.
\end{abstract}


\IEEEpeerreviewmaketitle

\section{Introduction}

\label{sec:Introduction}
Our research group proposed an estimation method to calculate the agent's emotion from the contents of utterances and to express emotions which are aroused in computer agent by using synthesized facial expression \cite{Ichimura03, Mera02, Mera03}. Emotion Generating Calculations (EGC) method \cite{Mera03} based on the Emotion Eliciting Condition Theory \cite{Elliott92} can decide whether an event arouses pleasure or not and quantify the degree of pleasure under the event.

 The calculated emotion by EGC is changed according to the emotions of the agent. Ren \cite{Ren06} describes Mental State Transition Network (MSTN) which is the basic concept of approximating to human psychological and mental responses. The assumption of discrete emotion state is that human emotion is classified into some kinds of stable discrete states, called ``mental state,'' and the variance of emotions occurs in the transition from a state to the other state with an arbitrary probability. Mera and Ichimura \cite{Mera10, Ichimura13} developed a computer agent that can transit a mental state in MSTN to the other state according to the kinds of emotion generated by EGC. The strength of emotion and the type of the aroused emotion by EGC arouses the transition of mental states\cite{Mera10}. MSTN can measure the user's feeling which is the mood represented in the continuous transition of mental states. 

 We have developed the Android EGC application software which the agent works to evaluate the feelings in the conversation\cite{Ichimura12a}. The smartphone user can not only obtain the variety of information but also converse with the agent in a smartphone, because the interface between human and smartphone has been equipped with the speech recognition. Our proposed techniques, EGC and MSTN, can be expected to be an emotional orientated intelligent interface.

 The application using Android EGC enables the estimation of user's feelings. We developed the tourist information system which can estimate the user's feelings at the sightseeing spot. The system can recommend the sightseeing spot and the local food corresponded to the user's feeling. The current recommendation systems nominate many information including the matter that the user thinks back the past sad occurrence, even if he/she feels unutterable solitude and desolation. By such recommendation, a traveler will stop the sightseeing on the way. 

Our developed system for Hiroshima Tourist can guide some spots, local food shops, and local gifts collected in Hiroshima Tourist Map Android application software\cite{Android_Market}. Because the smartphone has the GPS device and acceleration sensor, the application system has a navigation system. In this paper, the system decides the next candidates for spots and foods according to the tourist feelings by EGC in order to enjoy the travel. The system calculates the recommendation list by the estimate function of which the number of hits for a term retrieved by Google search, the important degree of a term included in Hiroshima sightseeing website, and the strength of emotion and the type of the aroused emotion by EGC. In order to show the effectiveness, this paper describes the experimental results for some situations during Hiroshima sightseeing.

The remainder of this paper is organized as follows. In the section \ref{sec:EGC}, the brief explanation to understand the EGC is described. Section \ref{sec:Recommendation_Tourist} describes the proposed recommendation method which combines 3 following methods; Google search results, the important degree of a term at the sightseeing website, and the the aroused emotion by EGC. In Section \ref{sec:Experimental_Results}, the recommendation system is explained. In Section \ref{sec:ConclusiveDiscussion}, we give some conclusive discussions that carries out practical experiments to cooperate with local government.

\section{Emotion Generating Calculations}
\label{sec:EGC}
\subsection{An Overview of Emotion Generating Process}

\begin{figure}[ht]
\begin{center}
\includegraphics[scale=0.8]{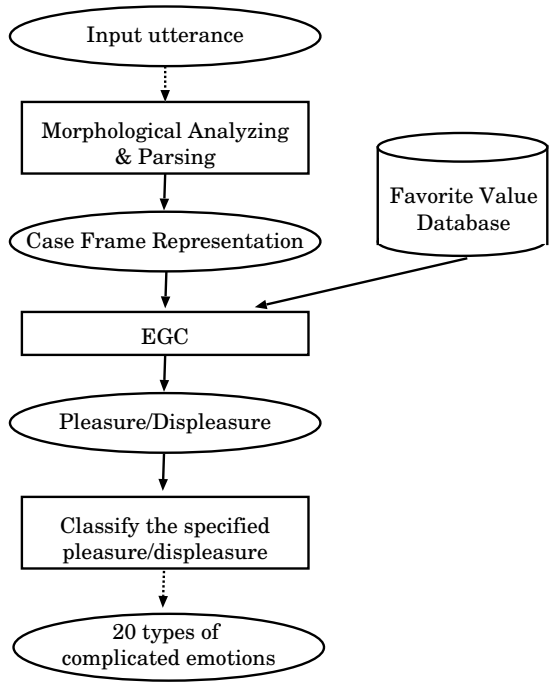}
\caption{Process for generating emotions}
\label{fig:processgeneratingemotion}
\end{center}
\end{figure}

Fig.\ref{fig:processgeneratingemotion} shows the emotion generating process where the user's utterance is transcribed into a case frame representation based on the results of morphological analysis and parsing. The agent works to determine the degree of pleasure/displeasure from the event in case frame representation by using EGC. EGC consists of 2 or 3 terms such as  subject, object and predicate, which have Favorite Value ($FV$), the strength of the feelings described in section \ref{sec:FavoriteValue}.

Then, the agent divides this simple emotion (pleasure/displeasure) into 20 various emotions based on the Elliott's ``Emotion Eliciting Condition Theory\cite{Elliott92}.'' Elliott's theory requires judging conditions such as ``feeling for another,'' ``prospect and confirmation,'' and ``approval/disapproval.'' The detail of this classification method is described in the section \ref{sec:ComplicatedEmotion}.

\subsection{Case Frame Representation}
\begin{figure}[!ht]
\begin{center}
\includegraphics[scale=1.0]{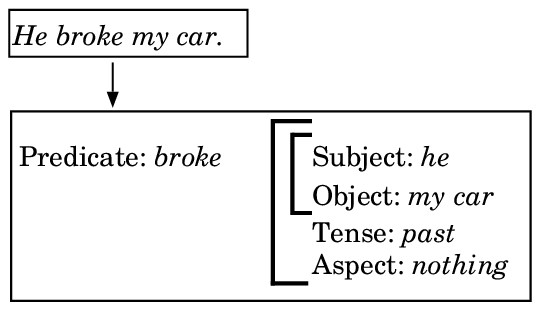}
\caption{Example of Case Frame Structure}
\label{fig:caseframerepresentation}
\end{center}
\end{figure}

The case frame structure bases the predicate phrase and the syntactic dependency between it and the other case elements. Fillmore\cite{Fillmore68} developed a system of linguistic analysis where the theory analyzes the surface syntactic structure of sentences by the combination of deep cases, e.g. semantic roles. Each verb selects a certain number of deep cases which form its case frame as shown in Fig.\ref{fig:caseframerepresentation}. 

In order to transcribe the user's utterances into the case frame representation, we implement morphological analysis and parsing to the input sentence.

\subsection{Favorite Value Database}
\label{sec:FavoriteValue}
Which an event is pleasure or displeasure is determined by using $FV$. $FV$ is a positive/negative number to an object when the user likes/dislikes it, respectively. $FV$ is predefined a real number in the range $[-1.0, 1.0]$. There are two types of $FV$s, personal $FV$ and initial $FV$. Personal $FV$ is stored in a personal database for each person who the agent knows well, and it shows the degree of like/dislike to an object from the person's viewpoint. On the other hand, an initial $FV$ shows the common degree of like/dislike to an object that the agent feels. Generally, it is generated based on the agent's own preference information according to the result of some questionnaires. Both personal and initial $FV$s are stored in the user own database. An initial value of $FV$ is determined beforehand on the basis of `corpus' of its applied field. The $FV$s of the objects are gained from a questionnaire. However, there are countless objects in the world. In this paper, we limit the objects that have initial $FV$ into the frequently appeared words in the dialog during sightseeing.

\subsection{Equation of EGC}
We assume an emotional space as three-dimensional space. Therefore, we present a method to distinguish pleasure/displeasure from an event by judging the existence of `synthetic vector''\cite{Mera02}.

\begin{table}[tbp]
\begin{center}
\caption{Correspondence between the event type and the axis}
\begin{tabular}{c|c|c|c}
\hline
Event type  & $f_{1}$ & $f_{2}$ & $f_{3}$ \\ \hline 
$V(S)$      &         &        &         \\
$A(S,C)$    &         &        &         \\
$A(S,OF,C)$ &         &        &         \\
$A(S,OT,C)$ & $f_{S}$  &        & $f_{P}$ \\
$A(S,OM,C)$ &         &        &         \\
$A(S,OS,C)$ &         &        &         \\ \hline
$V(S,OF)$   & $f_{S}$ & $f_{OT}-f_{OF}$ & $f_{P}$  \\
$V(S,OT)$   &         &         &         \\ \hline
$V(S,OM)$   & $f_{S}$ & $f_{OM}$ & $f_{P}$  \\ \hline
$V(S,OS)$   & $f_{S}-f_{OS}$ &  & $f_{P}$  \\ \hline
$V(S,O)$    & $f_{S}$ & $f_{O}$ & $f_{P}$  \\ 
            & $f_{O}$ &         & $f_{P}$ \\ \hline
$V(S,O,OF)$ & $f_{O}$ & $f_{OT}-f_{OF}$ & $f_{P}$ \\
$V(S,O,OT)$ &         & $f_{OM}$       &  \\ \hline
$V(S,O,OM)$ & $f_{O}$ & $f_{OM}$ & $f_{P}$ \\ \hline
$V(S,O,I)$ & $f_{O}$ & $\mid f_{I} \mid$  & $f_{P}$ \\ \hline
$V(S,O,OC)$ & $f_{O}$ &          & $f_{OC}$ \\ \hline
$A(S,O,C)$  & $f_{O}$ &          & $f_{P}$ \\ \hline
\end{tabular}
\label{tab:EGC-eventtype}
\end{center}
\end{table}

Table \ref{tab:EGC-eventtype} shows the correspondence between the case element in EGC equations and the axis in the three-dimensional model. In Table \ref{tab:EGC-eventtype}, `V(S,*)' is the type of event (verb) and `A(S,*)' is the type of attribute (adjective). the variables denoted in Table \ref{tab:EGC-eventtype} are expressed as follows.

\begin{itemize}
\item $f_{S}$ : $FV$ of Subject
\item $f_{OF}$ : $FV$ of Object-From
\item $f_{OM}$ : $FV$ of Object-Mutual
\item $f_{OC}$ : $FV$ of Object-Content
\item $f_{O}$ : $FV$ of Object
\item $f_{OT}$ : $FV$ of Object-To
\item $f_{OS}$ : $FV$ of Object-Source
\item $f_{P}$ : $FV$ of Predicate
\item $f_{I}$ : $FV$ of Instrument or tool
\end{itemize}

\begin{figure}[btp]
\begin{center}
\includegraphics[scale=0.5]{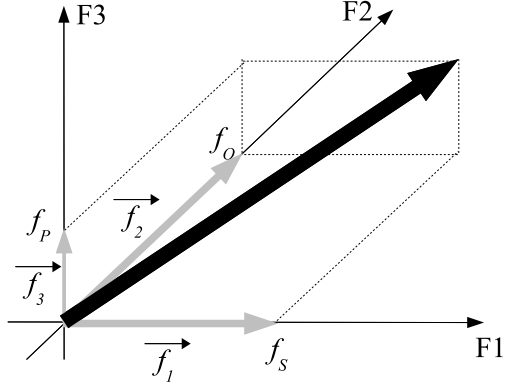}
\caption{Emotion Space for EGC}
\label{fig:EGC-emotionvector}
\end{center}
\end{figure}

Table \ref{tab:EGC-axis} shows the relation between the sign of axis in each dimension and the pleasure/displeasure of generated emotion. When the vector is on the axis, the event does not arouse any emotion. When we calculate the synthetic vectors of the events which do not have $f_{i}$ elements, we supply a dummy $FV$, $\beta$ as $f_{i}$ element. We tentatively defined $\beta$ as $+0.5$. Fig.\ref{fig:EGC-emotionvector} is an example of emotion space of event type $V(S, O)$. There are three elements, Subject, Object, and Predicate, in the event type, and the orthogonal vectors by the elements construct a rectangular solid. 

\begin{table}[tbp]
\begin{center}
\caption{pleasure/displeasure in emotional space}
\begin{tabular}{c|c|c|c|c}
\hline
Area & $f_{1}$ & $f_{2}$ & $f_{3}$ & Emotion \\ \hline
I & + & + & + & Pleasure \\
I\hspace{-.1em}I & - & + & + & Displeasure \\
I\hspace{-.1em}I\hspace{-.1em}I & - & - & + & Pleasure \\
I\hspace{-.1em}V & + & - & + & Displeasure \\
V & + & + & - & Displeasure \\
V\hspace{-.1em}I & - & + & - & Pleasure \\
V\hspace{-.1em}I\hspace{-.1em}I & - & - & - & Displeasure \\
V\hspace{-.1em}I\hspace{-.1em}I\hspace{-.1em}I & + & - & - & Pleasure \\ \hline
\end{tabular}
\label{tab:EGC-axis}
\end{center}
\end{table}

\subsection{Complicated Emotion Eliciting Method}
\label{sec:ComplicatedEmotion}
\begin{figure}[btp]
\begin{center}
\includegraphics[scale=0.9]{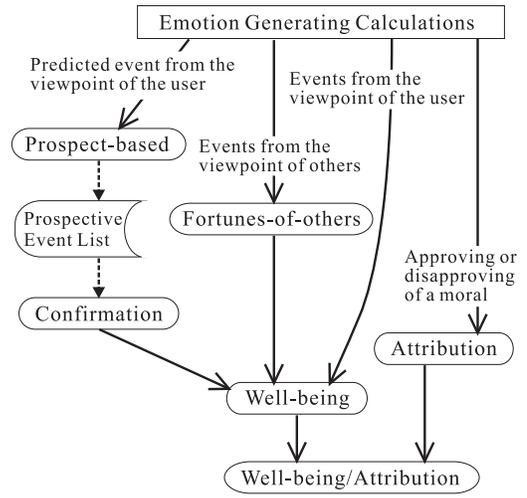}
\caption{Dependency among emotion groups}
\label{fig:EGC}
\end{center}
\end{figure}

Based on emotion values calculated by EGC method and their situations, the pleasure/displeasure is classified into 20 types of emotion. We consider only 20 emotion types, which are classified into six emotional groups as follows, ``joy'' and ``distress'' as a group of ``Well-Being,'' ``happy-for,'' ``gloating,'' ``resentment,'' and ``sorry-for'' as a group of ``Fortunes-of-Others,'' ``hope'' and ``fear'' as a group of ``Prospect-based,'' ``satisfaction,'' ``relief,'' ``fears-confirmed,'' and ``disappointment'' as a group of ``Confirmation,'' ``pride,'' ``admiration,'' ``shame,'' and ``disliking'' as a group of ``Attribution,'' ``gratitude,'' ``anger,'' ``gratification,'' and ``remorse'' as a group of ``Well-Being/Attribution'' \cite{Mera03}. Fig.\ref{fig:EGC} shows the dependency among the groups of emotion types. 

\section{Recommendation for Tourist}
\label{sec:Recommendation_Tourist}
Our developed system for Hiroshima Tourist can guide some spots, local food shops, and local gifts which were collected in Hiroshima Tourist Map Android application software\cite{Android_Market,Ichimura12b}. Because smartphone has the GPS device and acceleration sensor, the application system has a navigation system. In this paper, the system decides the candidates for spots and foods according to the tourist feelings by EGC in order to enjoy the travel. The system makes the recommendation list by the estimate function which consists of information retrieval by Google search, TF-IDF for Hiroshima tourism website, and the EGC results. The agreement values on the 3 scales are calculated for the words representing spots, food and gifts and are composed into one vector. In this paper, the 2,284 words for Hiroshima travel were extracted from the top 50 articles of July 4, 2013 in the blog site; 4travel( http://4travel.jp/ ). 

\subsection{Information retrieval by Google search}
\label{sec:GoogleSearch}
\begin{figure}[ht]
\begin{center}
\includegraphics[scale=0.35]{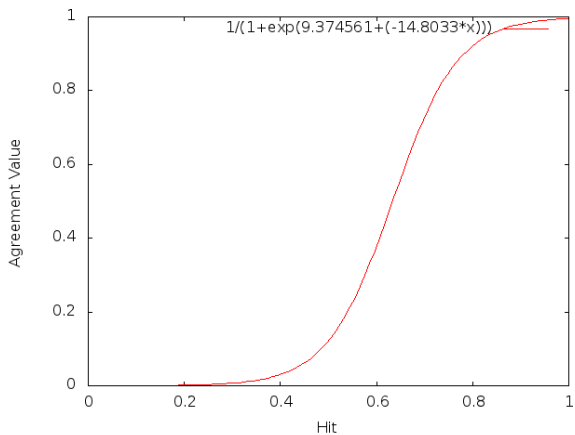}
\vspace{-4mm}
\caption{Estimation Function for Google Search}
\label{fig:GoogleSearch}
\end{center}
\end{figure}

 In Google search, the best page tends to be the ones that people linked to the most. Moreover, the best description of a page is often derived from the anchor text associated with the links to a page. The information technologies surrounding search engines is commonly referred to as information retrieval.
 The numerical number at the top of the page represents the retrieval results. The cumulative frequency distribution of the retrieval results for the 2,284 words was illustrated as shown in Fig.\ref{fig:GoogleSearch}. As a result, the estimation function for Google search is assumed as Eq.(\ref{eq:GoogleSearch}). In Eq.(\ref{eq:GoogleSearch}), $x$ means the number of retrieval results by Google search to the selected word. The $x$ axis and the $y$ axis of Fig.\ref{fig:GoogleSearch} show the normalized number of retrieval and the cumulative frequency number, respectively.
\begin{eqnarray}
\label{eq:GoogleSearch}
g_1(x) = \frac{1}{1+e^{9.374561+(-14.8033*x)}}
\end{eqnarray}

\subsection{TF-IDF for Hiroshima Tourism Web Information}
\label{sec:TF_IDF}
Next, we investigated the TF-IDF value of words in Hiroshima Tourism Web sites as shown in Table \ref{tab:Websitelist}.
\begin{eqnarray}
\nonumber tf(t, d)&=&\frac{n(t,d)}{\sum_{k} n(k,d)}\\
\nonumber idf(t) &=& \log \frac{|D|}{|\{d \in D: t \in d\}|}\\
tfidf(t,d)&=&tf(t,d) \times idf(t),
\label{eq:tfidf}
\end{eqnarray}
where $n(t,d)$ is the occurrence count of a term $t$ in the document $d$. $|D|$ is the total number of documents in the corpus. $|\{d \in D: t \in d\}|$ is the number of documents where the term $t$ appears.

In this paper, $d$ and $t$ indicate the html file in Tourist Web site and the representative word in the user comments of Android Application, respectively.

However, this simulation designs to use only one TF-IDF value per a sample. A user does not know the number of words representing tourism information in one article and does not determine the `best' TF-IDF value. If a sample has two or more words with TF-IDF value, it is difficult to select the representative word per a sample. Because tourist's subjective data in our developed MPPS (MobilePhone based Participatory Sensing System)\cite{Ichimura12b, Android_Market} relates to sightseeing spots in Hiroshima, the words used in comments are also limited to the familiar location or gifts.

 The collected samples include valuable information that is known to few people. If there is a word with the maximum value of TF-IDF and it is a representative value in a sample, there was no appreciable difference among other words because most of TF-IDF value is small. In this paper, we consider that each comment has $N$ words at most and they are divided to 2 groups: the existing $l$ words with higher TF-IDF value and the remaining $N-l$ words with lower TF-IDF value. The TF-IDF value for $N-l$ words is 0. The experiment describes TF-IDF values in case of `$l=1$,' because the words with high TF-IDF values are not so much. Then, the value of TF-IDF field as shown in Table 3 denotes the sum of TF-IDF value for $l$ words in a comment.

\begin{table}[!tb]
\caption{Tourism association website}
\vspace{-4mm}
\label{tab:Websitelist}
\begin{tabular}{llr}
\noalign{\smallskip}\hline\noalign{\smallskip}
WebSite & URL & Words\\
\noalign{\smallskip}\hline\noalign{\smallskip}
Hiroshima & {\it http://www.kankou.pref.hiroshima.jp/foreign/english/} & 4,138\\
Kure & {\it http://www.urban.ne.jp/home/kurecci/} & 864\\
Hatsukaichi & {\it http://www.hatsu-navi.jp/} & 121\\
Onomichi &  {\it http://www.ononavi.com/} & 6,212\\
Fukuyama & {\it http://www.fukuyama-kanko.com/event/} & 2,424\\
Miyoshi & {\it http://www.kankou-miyoshi.jp/} & 1,664\\
Akitakata & {\it http://www.akitakata.jp/kankou/} & 3,624\\
\noalign{\smallskip}\hline\noalign{\smallskip}
\end{tabular}
\end{table}

\begin{figure}[ht]
\begin{center}
\includegraphics[scale=0.35]{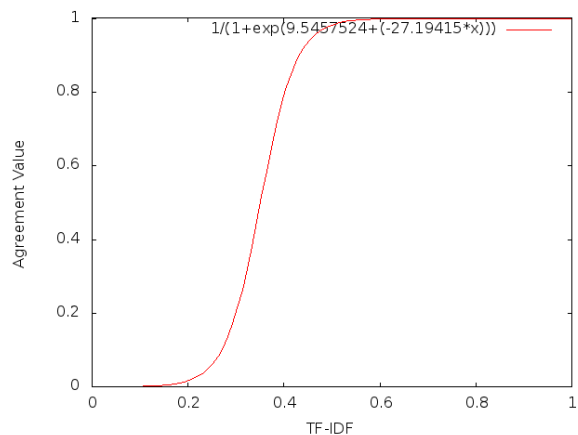}
\vspace{-4mm}
\caption{Estimation Function for TF-IDF Local Websites}
\label{fig:TF_IDF}
\end{center}
\end{figure}

The cumulative frequency distribution for TF-IDF values of 2,284 words was illustrated as shown in Fig.\ref{fig:TF_IDF}. As a result, the estimation function for TF-IDF is assumed as Eq.(\ref{eq:TF_IDF}). In Eq.(\ref{eq:TF_IDF}), $x$ means the TF-IDF value to the selected word. The $x$ axis and the $y$ axis of Fig.\ref{fig:TF_IDF} show the normalized number of retrieval results and the cumulative frequency number, respectively.

\begin{eqnarray}
\label{eq:TF_IDF}
g_2(x) = \frac{1}{1+e^{9.5457524+(-27.19415*x)}}
\end{eqnarray}

\subsection{Estimation by EGC}
\label{sec:EGC_Estimation}
EGC can calculate the type and its degree of user's emotion described in the section \ref{sec:ComplicatedEmotion}. In the system, we consider that the spots and foods with pleasure emotion by EGC will be recommended. On the contrary, if the user meets the situation where displeasure emotion will be aroused, the user feels more deep displeasure. The words arousing the user's displeasure were recorded in the taboo list and then the system avoids to use them.

\subsection{Total Estimation}
\label{sec:Total_Estimation}
The total estimation function for the recommendation is synthesized with the 3 dimension Eq.(\ref{eq:GoogleSearch}), Eq.(\ref{eq:TF_IDF}), and EGC results.
\begin{eqnarray}
\label{eq:total_estimation}
\nonumber Recommend(W_{i}) &=& \sqrt{\mathstrut (g_{1}(W_{i}))^2+(g_{2}(W_{i}))^2 + (EGC_{i})^2},\\
 i&=&1, \cdots, N, 
\end{eqnarray}
where $W_{i}$ means the word in the sentence $i$ and $EGC_{i}$ means the emotion value for the sentence $i$.

In this paper, the system recommends the place, the food, or the gift corresponding to a noun in the sentence of the user's utterance. In \cite{Ichimura12a}, we proposed the concierge system for Hiroshima Tourist by using Fuzzy Petri Net\cite{Chen91}. The user's utterances are classified into 2 main types according to the following verbs $\{`see', `go', `come'\}$ related to the action for place and $\{`eat,' `(be) hungry,' `buy,' `look up'\}$ related to the action for eating and purchasing.

 If the user's utterance includes the word representing the place, the system selects the top 10 places among the list of the recommended place near the place in the user's utterance after the system calculates the recommended value for the sentence by Eq.(\ref{eq:total_estimation}). The $Recommend(W_{i})$ at the places in the list becomes the higher than the place of the user's utterance, because the recommended places were enumerated in the order increasing evaluation value. 

On the contrary, if the user's utterance includes the word representing the food or gift, the system selects the top 5 foods or gifts among the list in similar way. The $Recommend(W_{i})$ for the food or the gift in the list becomes the higher than that of the user's utterance.

\section{Experimental Results}
\label{sec:Experimental_Results}

In this paper, we investigated what kind of the recommendation was performed for the following sentences under the situation in Hiroshima sightseeing.

\begin{enumerate}
\item I am going to get to the Hiroshima castle.
\item I would like to eat a lunch.
\end{enumerate}

\begin{figure}[ht]
\begin{center}
\includegraphics[scale=0.25]{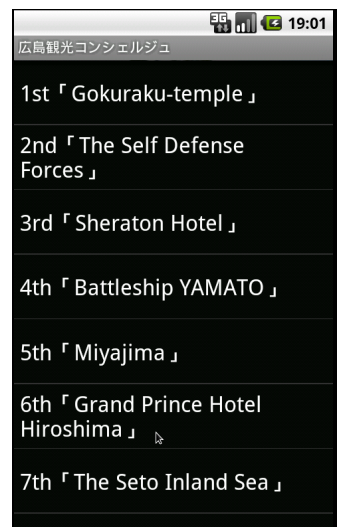}
\caption{Recommendation for spot}
\label{fig:recommendation_spot}
\end{center}
\end{figure}

First, the emotion value is calculated by EGC. For the sentence 1), 3 elements in caseframe representation are $f_{1}=0.5$, $f_{2}=0.0$, and $f_{3}=0.0$, respectively. As a result, the degree of pleasure is $0.2887$. That is, the emotion value is $0.2887$ and the type is `joy' and `happy-for'. Second, as mentioned in section \ref{sec:Total_Estimation}, the system recommended the sightseeing spots as shown in Table \ref{tab:recommendlist_spot}. The recommended values for the places in the list are equal to or higher than that of current utterance. The system shows the recommendation list as shown in Fig.\ref{fig:recommendation_spot}.

\begin{table}[!tb]
\begin{center}
\caption{Recommended list for sightseeing spots}
\label{tab:recommendlist_spot}
\tiny
\begin{tabular}{|c|c|c|c|c|c|}
\hline
&Word&$g_1(W_i)$&$g_2(W_i)$&$EGC$&$Rec(W_i)$\\ \hline
1&Gokuraku Temple& 0.0958& 0.8020&0.7071& 1.0735\\ \hline
2&Battleship Yamato& 0.7082& 5.1722E-4&0.7071& 1.0008\\ \hline
3&The Self Defense Forces& 0.7639& 2.9405E-4&0.5830& 0.9610\\ \hline
4&Miyajima& 0.3092& 7.9025E-4&0.8602& 0.9141 \\ \hline
5&Itukushima Shrine& 0.1335& 1.6933E-4& 0.8602&0.8705\\ \hline
6&Torii& 0.1251& 9.3482E-5&0.8602& 0.8693\\ \hline
7&Momijidani Park& 0.2499& 7.5539E-5&0.7071& 0.7500\\ \hline
8&Rihga Royal Hotel Hiroshima& 0.1705& 8.912E-5&0.7071& 0.7274\\ \hline
9&View the scarlet maple leaves& 0.1207& 1.0161E-4&0.7071& 0.7173 \\ \hline
10&Hiroshima Peace Memorial Park& 0.1136& 0.0025& 0.7071&0.7162\\
\hline
\end{tabular} 
\end{center}
\end{table}

\begin{figure}[ht]
\begin{center}
\includegraphics[scale=0.20]{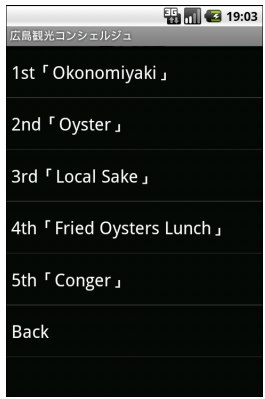}
\caption{Recommendation for food}
\label{fig:recommendation_food}
\end{center}
\end{figure}

\begin{figure}[ht]
\begin{center}
\includegraphics[scale=0.25]{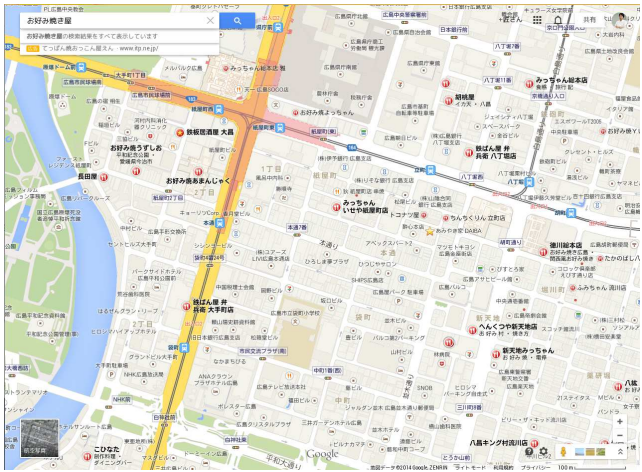}
\caption{Restaurants on the Google map}
\label{fig:GoogleMap}
\end{center}
\end{figure}

For the sentence 2), 3 elements in caseframe representation are $f_{1}=0.0$, $f_{2}=0.6$, and $f_{3}=0.0$, respectively. As a result, the degree of pleasure is $0.3464$. That is, the emotion value is $0.3464$ and the type is `joy' and `happy-for'. Second, as mentioned in section \ref{sec:Total_Estimation}, the system recommended the local food category as shown in Table \ref{tab:recommendlist_food}. The system shows the recommendation list as shown in Fig. \ref{fig:recommendation_food}. If the user clicks one in the list, the system opens the Google map on the user's current location as shown in Fig. \ref{fig:GoogleMap}. The recommended values for the categories in the list are  equal to or higher than that of current utterance. Next, the user can use the Navigation application to get the target restaurant. 

\begin{table}[htbp]
\begin{center}
\caption{Recommended list for food}
\label{tab:recommendlist_food}
\begin{tabular}{|c|c|c|c|c|c|}
\hline
&Word&$g_1(W_i)$&$g_2(W_i)$&$EGC$&$Rec(W_i)$\\ \hline
1&Okonomiyaki& 0.5563& 0.0070&1.1912& 1.3115\\ \hline
2&Fried Oysters Lunch& 0.2937& 7.3239E-5&1.0488& 1.0892\\ \hline
3&Conger& 0.2547& 7.6289E-5&1.0488& 1.0793\\ \hline
4&Oyster& 0.5022& 0.0022&0.9273& 1.0546 \\ \hline
5&Local Sake& 0.3622& 7.5449E-5& 0.9273&0.9956\\
\hline
\end{tabular} 
\end{center}
\end{table}

\begin{figure}[tb]
\begin{center}
\includegraphics[scale=0.28]{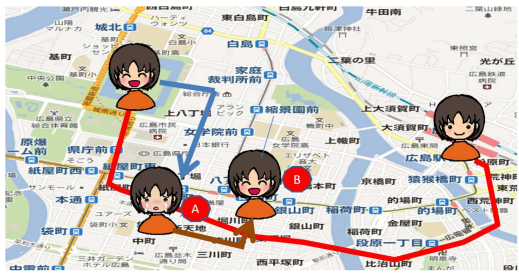}
\caption{Hiroshima Tourist Information Concierge System}
\label{fig:concierge}
\end{center}
\end{figure}

Fig.\ref{fig:concierge} shows the simulation result to act as guide from Hiroshima station to the downtown. The user starts from Hiroshima station in the right side in Fig.\ref{fig:concierge} and goes to the downtown in the center in Fig.\ref{fig:concierge}. Then the user reaches the first destination, Hiroshima Castle. Next, after the user talks with the system, the concierge recommends `Okonomi-Yaki' restaurant. The system can recognize the user's emotion value and type by EGC and then recommend the favorite local food, `Okonomi-Yaki'. The user selects the category and the Google map shows the restaurant $\textcircled{\footnotesize A}$ near the current place as shown in Fig.\ref{fig:concierge}. The user went to the $\textcircled{\footnotesize A}$, however, the restaurant was crowded and the user could not enter. The system recommended another restaurant $\textcircled{\footnotesize B}$ according to this situation. The user can enjoy a comfortable sightseeing by the end of the day.

\section{Conclusion}
\label{sec:ConclusiveDiscussion}
The smartphone can use various kinds of applications such as web browser, e-mail, Google map and so on. Especially, the voice recognition function is the outstanding application to spread the capability of mobile phone, because the current dialog system requires the user's typing. For example, the concierge system uses voice recognition function. However, the essential quality of dialog is limited to question and answer, although the recognition rate becomes good. In order to enjoy real conversation, the system can evaluate the user's emotion by using Android EGC\cite{Ichimura12a}. However, the recommendation system in this paper does not work with the MSTN, because MSTN can represent mood transition for aroused emotions. In order to reply different response for same input and different mood, we will develop the response text database for each mood.

The Hiroshima tourist website replenishes the variety of information for foreigners. The Android application `Hiroshima Tourist map'\cite{Android_Market} has been developed to one of Mobile Phone based Participatory Sensing system, because not only tourism association but the local citizens should give the innovative and attractive information in sightseeing to visitors. We will embed the concierge system into our developed `Hiroshima Sightseeing map' in near future. The usability for the developed system will be investigated to put the system in practical use such as the development of special regional products, tourism resources, and markets in Etajima City.

\section*{Acknowledgment}
This work was supported by JSPS KAKENHI Grant Number 25330366.

\end{document}